# ACOUSTO-OPTIC FREQUENCY MODULATOR

*S. Tallur[*], and S.A. Bhave*
OxideMEMS Lab, Cornell University, Ithaca, New York, USA

## ABSTRACT

Frequency modulation of a continuous wave laser source enables generation of photons at multiple frequencies from a single pump laser. Such multiple closely spaced laser lines are important for high data rate QAM in dense-WDM networks. In this paper we present a monolithic silicon acousto-optic frequency modulator (AOFM) operating at 1.09GHz. We demonstrate frequency modulation of a 1564nm wavelength pump laser, resulting in generation of sideband laser lines spaced by 0.009nm.

## INTRODUCTION

On-chip optical frequency modulation has been demonstrated by coupling surface acoustic waves (SAW) to InAs/GaAs quantum dots [1, 2]. This phenomenon has not been demonstrated in silicon to date. The silicon acousto-optic modulator (AOM) presented in [3] has a mechanical resonance frequency of 1.1GHz but an optical cavity 3dB-linewidth ≈5.88GHz. The mechanical response of the resonator is hence slower than the lifetime of the photons inside the cavity and thus results in predominant amplitude modulation (AM) of the laser light. Here we present a modulator with an optical cavity linewidth ≈ 1.27GHz. The rate of the mechanical vibrations is comparable to the photon lifetime, which results in Doppler-shift of the circulating intra-cavity optical field, thus causing frequency modulation (FM). Such an Acousto-Optic Frequency Modulator (AOFM) converts photons at the pump laser frequency to other frequencies on account of mechanical motion.

When the mechanical vibration frequency is greater than half of the optical cavity 3dB-linewidth, the resonator is termed to lie in the "resolved sideband regime" [4]. In this case, the motion of the optomechanical resonator results in asymmetric higher and lower frequency optical sidebands spaced by the mechanical resonance frequency. The asymmetry can be explained solely by coincident AM and FM [5]. For the modulator presented here, the lower frequency sideband has >12X higher intensity than the higher frequency sideband at 1.09GHz, indicating strong FM plus some AM.

## DEVICE DESIGN

When the mechanical resonance frequency for the modulator, $\Omega_m$, is greater than half of the optical cavity linewidth $\kappa/2 = \omega_{opt}/2Q_{opt}$, the dependence of the intra-cavity field amplitude on the mechanical frequency is complex [6] and one can write the exact solution [4]:

$$a_p(t) = s\sqrt{\kappa} \sum_{n=-\infty}^{n=+\infty} \frac{(-i)^n J_n(\beta)}{\frac{\kappa}{2} + i(\Delta + n\Omega_m)} e^{i(\omega + n\Omega_m)t + i\beta\cos(\Omega_m t)} \quad (1)$$

The modulation index, $\beta$, depends linearly on the amplitude of mechanical displacement of the micro-ring. The offset of the pump laser frequency from the cavity resonance frequency is denoted by the detuning, $\Delta = \omega_{laser} - \omega_{opt}$. The electro-acoustically driven motion of the cavity leads to a combination of AM and FM of the laser light as seen in equation 1.

The modulator presented in [3] is extended to an array of mechanically coupled ring resonators for large electrostatic driving force [7]. Figure 1 shows a scanning electron micrograph (SEM) of the device. Mechanical motion of the ring resonators is actuated through capacitive air gap electrostatic transduction. The lithographically defined air gap is 130nm. Input laser light is coupled into the optomechanical resonator using an integrated waveguide.

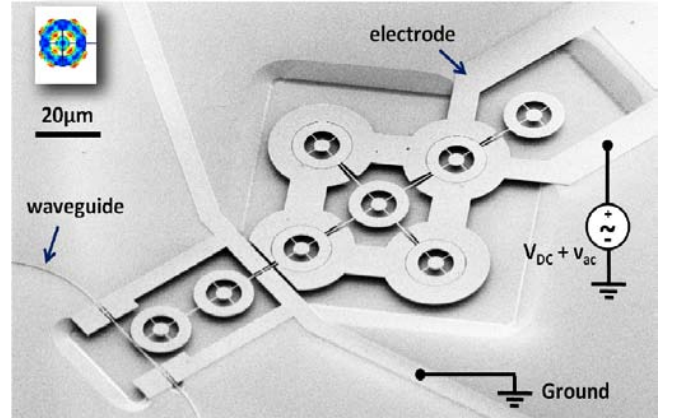

*Figure 1: SEM of the acousto-optic frequency modulator. (Top-left inset: COMSOL mode shape of the compound radial expansion mode of the ring resonator at 1.09GHz)*

## EXPERIMENTAL RESULTS AND DISCUSSION

To probe the optical response of the modulator, light from a Santec TSL-510 tunable laser is coupled into the waveguide using grating couplers. We choose an optical resonant mode at a wavelength of 1564nm with an optical quality factor of 150,000, which corresponds to a cavity linewidth ≈1.27GHz. The laser wavelength is blue detuned to the half maximum point of this optical cavity mode and the transmission of the modulator is obtained by performing a 2-port measurement using an Agilent N5230A network analyzer. A mechanical quality factor of 1,800 was measured for the resonant mode at 1.09GHz as shown in Figure 2.

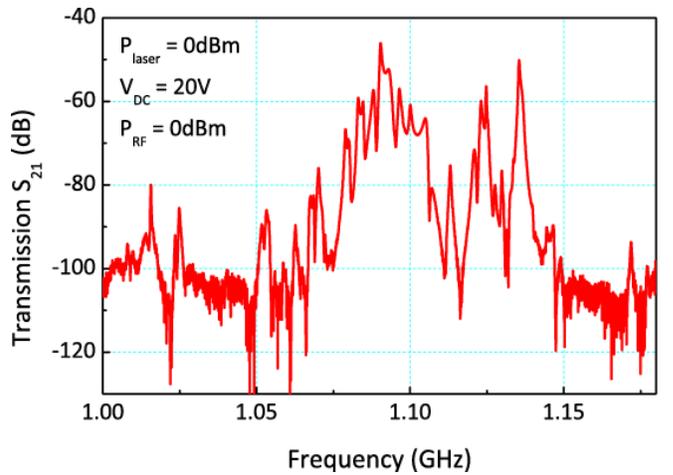

*Figure 2: RF transmission of the AOFM. Multiple peaks corresponding to different modes of vibration of the device are seen. The mechanical resonant mode at 1.09GHz that corresponds to in phase motion of all the rings in the array has an insertion loss of 46dB.*

For observing the optical sidebands for the modulator, we use a Thorlabs SA210-12B scanning Fabry-Perot (FP) interferometer. A bias-tee is used to apply a combination of DC bias voltage and AC voltage at 1.09GHz using an Agilent E8257D PSG Analog Signal Generator. The wavelength of the FP cavity is scanned across its entire range by sweeping the voltage on the piezo-controller in the FP cavity control box. The transmitted light intensity is measured using an internal photodiode, amplified by a transimpedance amplifier inside the control box, and displayed on an oscilloscope. An illustration of the setup is shown in Figure 3.

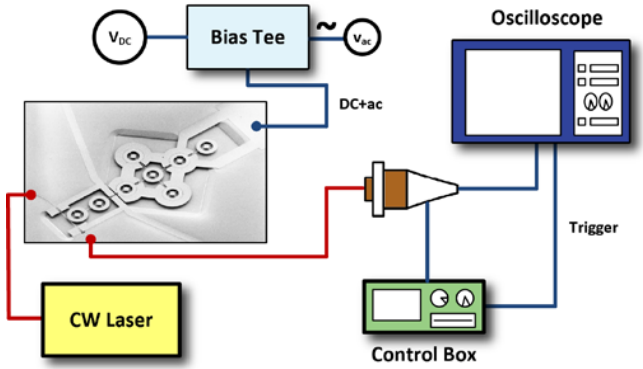

*Figure 3: Experimental setup for characterization of the optical sidebands of the AOFM using a scanning FP interferometer. As the FP cavity scans across wavelengths, its output shows a peak on the oscilloscope whenever it passes across a sideband.*

The optical sidebands are analyzed for relative detuning ($2\Delta/\kappa$) values of 0.5 and 1.5 respectively. Figure 4 shows the measured intensity values at the sidebands, normalized to the intensity at the input laser frequency incident on the FP interferometer. A combination of 20V DC and 8dBm RF power was applied at the input electrode of the modulator. The observed asymmetry in the sideband intensities results from coincidental AM and FM [5]. This is in sharp contrast to perfectly symmetric sideband generation on account of pure AM or pure FM.

The lower frequency sideband is enhanced by the optical cavity, while the higher frequency sideband is filtered out. This effect is maximized for $\Delta = \Omega_m$, which corresponds to a relative detuning of 1.5 (Figure 4(b)). For this case, the ratio of the sideband heights is 12.67:1, which compares well to the calculated value of 10.8 from simulation following equation 1. In comparison, we measure a ratio of 2.2X for relative detuning of 0.5 (Figure 4(a)). The modulation index, $\beta$, which appears in equation 1, was calculated by calculating the amplitude of mechanical displacement of the micro-ring by following the derivation in [8]. Although our final goal is pure FM of laser light, the asymmetry due to presence of some AM serves as a proof of concept.

## CONCLUSION

We demonstrate frequency modulation of a continuous wave laser of wavelength 1564nm using a silicon acousto-optic modulator operating at 1.09GHz, thus resulting in generation of sideband laser lines spaced apart by 9.23pm. This constitutes the first demonstration of electro-acoustically generating photons of different colors from a single pump laser in silicon.

## REFERENCES

[1] M. Metcalfe, et al., "Resolved sideband emission of InAs-GaAs quantum dots strained by surface acoustic waves", Phys. Rev. Lett. 105, 037401 (2010).

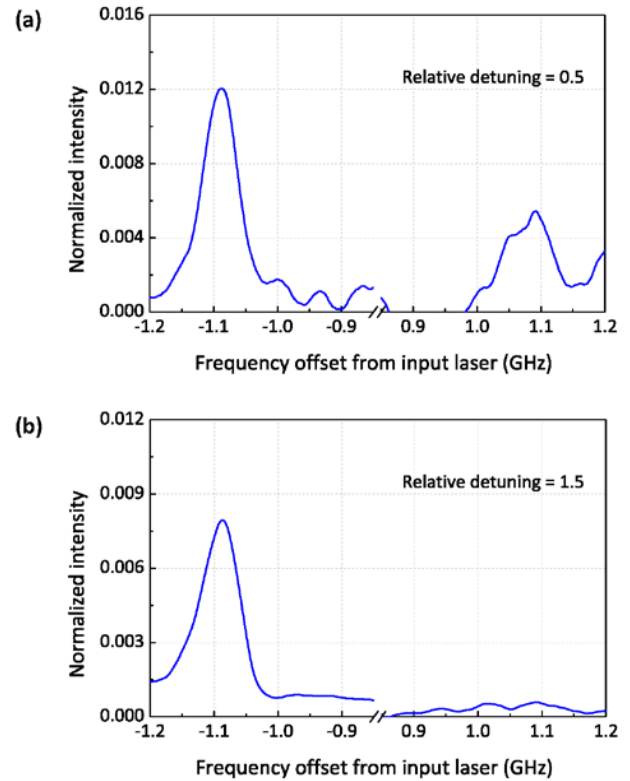

*Figure 4: Motional sidebands at ±1.09GHz offsets from laser frequency measured for relative detuning value of (a) 0.5, and (b) 1.5. In panel (b), the peak intensities at the sidebands have a ratio of 12.67:1. The asymmetry of the sidebands confirms simultaneous AM and FM.*

[2] D.A. Fuhrmann, et al., "Dynamic modulation of photonic crystal nanocavities using gigahertz acoustic phonons", Nature Photonics, 5, pp. 605-609 (2011).

[3] S. Sridaran and S. A. Bhave, "1.12 GHz opto-acoustic oscillator", 25th IEEE International Conference on Micro Electro Mechanical Systems, Paris, France (2012) pp 644-647.

[4] A. Schliesser, et al., "Resolved-sideband cooling of a micromechanical oscillator", Nature Physics 4, 415-419 (2008).

[5] H. Roder, "Amplitude, Phase, and Frequency Modulation", Proceedings of the Institute of Radio Engineers, 19(12), pp. 2145-2176 (1931).

[6] T. Carmon, et al., "Temporal Behavior of Radiation-Pressure-Induced Vibrations of an Optical Microcavity Phonon Mode", Phys. Rev. Lett. 94, 223902 (2005).

[7] S.-S. Li, Y.-W. Lin, Z. Ren, and C. T.-C. Nguyen, "An MSI micromechanical differential disk-array filter", 14[th] Int. Conf. on Solid-State Sensors & Actuators (Transducers'07), Lyon, France (2007) pp. 307-311.

[8] S. Tallur, T. J. Cheng, S. Sridaran and S. A. Bhave, "Motional impedance analysis: bridging the 'gap' in dielectric transduction", 65[th] IEEE Frequency Control Symposium, San Francisco, California (2011) pp. 135-138.

**CONTACT**
*S. Tallur, tel: +1-607-279-7818; sgt28@cornell.edu